\documentclass[twocolumn]{el-author}

\usepackage{algorithmic}

\makeatletter
\def\Ginclude@eps#1{%
 \message{<#1>}%
 \bgroup
 \def\@tempa{!}%
 \dimen@\Gin@req@width
 \dimen@ii.1bp%
 \divide\dimen@\dimen@ii
 \@tempdima\Gin@req@height
 \divide\@tempdima\dimen@ii
 \includegraphics{#1}%
 \egroup}
\makeatother

\newcommand{\ua}{\uparrow}
\newcommand{\nc}{\newcommand}
\nc{\da}{\downarrow} \nc{\hc}{\hat{c}} \nc{\hS}{\hat{S}}
\nc{\bra}{\langle} \nc{\ket}{\rangle} \nc{\eq}{equation (\ref}
\nc{\h}{\hat} \nc{\hT}{\h{T}}\nc{\be}{\begin{eqnarray}}
\nc{\ee}{\end{eqnarray}}\nc{\rd}{\textrm{d}}\nc{\e}{eqnarray}\nc{\hR}{\hat{R}}\nc{\Tr}{\mathrm{Tr}}
\nc{\tS}{\tilde{S}}\nc{\tr}{\mathrm{tr}}\nc{\8}{\infty}\nc{\lgs}{\bra\ua,\phi|}\nc{\rgs}{|\ua,\phi\ket}
\nc{\hU}{\hat{U}}\nc{\lfs}{\bra\phi|}\nc{\rfs}{|\phi\ket}\nc{\hZ}{\hat{Z}}\nc{\hd}{\hat{d}}\nc{\mD}{\mathcal{D}}
\nc{\bd}{\bar{d}}\nc{\bc}{\bar{c}}\nc{\mc}{\mathcal}\nc{\ea}{eqnarray}\nc{\mG}{\mathcal{G}}\nc{\bce}{\begin{center}}
\nc{\ece}{\end{center}}
\date{\today}

\begin{document}

\title{Diffusion Least Mean P-Power Algorithms for Distributed Estimation in Alpha-Stable Noise Environments}

\author{F. Wen}

\abstract{We propose a diffusion least mean p-power (LMP) algorithm for distributed estimation in alpha stable noise environments, which is one of the widely used models that appears in various environments. Compared with the diffusion least mean squares (LMS) algorithm, better performance is obtained for the diffusion LMP methods when the noise is with alpha-stable distribution.}

\maketitle

\section{Introduction}

Emergent wireless sensor networks based applications have motivated the development of distributed adaptive estimation schemes. Distributed least mean squares (LMS) \cite{Cattivelli2010} and recursive least squares (RLS) type algorithms have received more attentions \cite{Cattivelli2008}. 
Readers can refer to \cite{Sayed2013} and the references therein for details about up to date diffusion strategies for adaptation and learning over networks.

The distributed LMS and RLS type strategies are all second order statistics (SOS) based, 
the target is to minimize the mean square error (MSE) of the estimator. 
For the MSE based techniques, the noise is often assumed to
be white Gaussian with finite second-order statistics.
However, in some circumstances, the noise may not
have finite SOS, such as impulsive noise, which can be modeled by a heavy-tailed alpha stable distribution \cite{Shao1993}.

Alpha stable signal processing techniques have received more attentions. 
Parameter estimation and blind channel identification in alpha stable signal environments are introduced in \cite{Ma1995}. System identification problem in alpha stable noise environments is discussed in \cite{Weng2005}.
Particle filtering for acoustic source tracking in impulsive noise with alpha-stable process is proposed in \cite{Zhong2013}.

The existing LMP algorithms are all global or centralized, distributed LMP have not been studied yet.
In this paper, we exploit the diffusion LMP strategies for distributed estimation in impulsive noise environments. The additive impulsive noise is with symmetric alpha-stable distribution.

\section{Problem Formulation}
\label{Sec2}

Consider a set of $N$ nodes distributed over some geographic
region. At every time instant $n$, every node $k$
takes a scalar measurement $d_{k,n}$ and a regression column vector $\boldsymbol{u}_{k,n}$, which is correlated with $d_{k,n}$, and 
\begin{equation}\label{eq1}
d_{k,n} =  \boldsymbol{\omega}_o^T  \boldsymbol{u}_{k,n} + v_{k,n}
\end{equation}
where $v_{k,n}$ denotes the measurement of model noise and $T$ denotes transposition.
The objective is for every node in the network to use the data $\left\{ d_{k,n}, \boldsymbol{u}_{k,n} \right\}$
to estimate the unknown column vector $\boldsymbol{\omega}_o$. We assume that all the signals are real, and extension to complex case is straightforward.

Here we give a brief introduction about $\alpha$-stable distribution.
The characteristic function of $\alpha$-stable process \cite{Shao1993}  is described as:
\begin{equation}\label{eq38}
f(t) = \exp \big\{ j \delta t - \gamma |t|^{\alpha} \left[ 1 + j \beta \textrm{sgn}(t)\mathcal{S}(t,\alpha)\right]\big\}
\end{equation}
where
\begin{equation}\label{eq39}
\mathcal{S}(t,\alpha) = \begin{cases} \tan \left( \frac{\alpha\pi}{2} \right) &\mbox{if } \alpha \neq 1 \\ 
\frac{2}{\pi}\log|t| & \mbox{if } \alpha = 1.\end{cases}
\end{equation}

Here $\alpha \in (0,2]$ is the characteristic exponent of the alpha stable distribution. It measures the tails heaviness of the distribution. 
The smaller $\alpha$ is, the heavier tails the process is.
 In addition, $ -\infty < \delta < \infty$ is the location
parameter of the distribution, and $\beta \in [-1, 1]$ is the symmetry parameter. $\gamma > 0$ is the dispersion, which plays a role similar to the variance of the Gaussian distribution.  
The distribution is symmetric about its location parameter $\delta$ when $\beta = 0$. Such a distribution is called a symmetric alpha stable distribution (S$\alpha$S). Throughout the paper, we assume that the alpha stable noise is symmetric $\beta = 0$ and the location parameter $\delta = 0$. 

\section{Proposed Method}
\label{Sec3}
For global LMP, 
the parameter $\boldsymbol{\omega}$ is estimated by minimizing the following global cost function:
\begin{equation}\label{eq2}
\mathcal{J}^{global}_k(\boldsymbol{\omega}) = \sum_{k = 1}^{N}  \textrm{E} \Big\{ \left |d_{k,n} - \boldsymbol{w}^T\boldsymbol{u}_{k,n} \right |^p \Big\},
\end{equation}
where $\textrm{E} \{ \cdot \}$ is expectation operator.
The unknown parameters are updated along the steepest descent of the cost function in (\ref{eq2}), which is given by
\begin{equation}\label{eq3}
\boldsymbol{\omega}_n  = \boldsymbol{\omega}_{n-1} + \sum_{k = 1}^{N} \mu_k \left( \left| e_{k,n}\right|^{p-2} e_{k,n} \boldsymbol{u}_{k,n} \right),
\end{equation}
where $e_{k,n} = d_{k,n} - \boldsymbol{w}^T_{k,n}\boldsymbol{u}_{k,n}$ is the error signal.

For local estimation, the following cost function is considered, 
\begin{equation}\label{eq4}
\mathcal{J}^{local}_k({\omega}) = \sum_{l \in \mathcal{N}_k} c_{kl} \textrm{E}\Big\{\left |d_{l,n} - \boldsymbol{w}^T_{l,n}\boldsymbol{u}_{k,n}\right |^p\Big\}
\end{equation}
where $\mathcal{N}_k$ denotes the neighborhood of an node $k$ and the coefficients $\{c_{kl}\}$ determine which nodes $l \in \mathcal{N}_k$ should share their measurements $\{d_{l,n}, \boldsymbol{u}_{l,n}\}$ with node $k$.

For each nodes, first performs intermediate estimates by the following combination,
\begin{equation}\label{eq9}
\boldsymbol{\phi}_{k,n-1} = \sum_{l \in \mathcal{N}_k} a_{1,lk} \boldsymbol{\omega}_{l,n-1},
\end{equation}
where the coefficients $\{ a_{1,lk} \}$ determine which nodes should share their intermediate estimates $\{\boldsymbol{\omega}_{l,n-1}\}$ with node $k$.
With all the intermediate estimates, the nodes update their estimates by
\begin{equation}\label{eq10}
\boldsymbol{\psi}_{k,n} = \boldsymbol{\phi}_{k,n-1} + \mu_k \sum_{l \in \mathcal{N}_k} c_{lk} \left| e_{l,n}\right|^{p-2} e_{l,n} \boldsymbol{u}_{l,n},
\end{equation}
After updating, the second combination is performed as
\begin{equation}\label{eq11}
\boldsymbol{\omega}_{k,n} = \sum_{l \in \mathcal{N}_k} a_{2,lk} \boldsymbol{\psi}_{l,n},
\end{equation}
where the coefficients $\{ a_{2,lk} \}$ determine which nodes should share their intermediate estimates $\{\boldsymbol{\psi}_{l,n}\}$ with node $k$.
The diffusion LMP algorithm is summarized in Algorithm \ref{GDLMP}.
\begin{algorithm}
Initialization:\\
Given non-negative real coefficients $\{a_{1,lk}, a_{2,lk}, c_{lk}\}.$\\
Start with $\boldsymbol{\omega}_{l,-1} =\boldsymbol{0}$ for all $l$.
\begin{algorithmic}[1]
\FOR{$n=0$ to $n_{\max}$}
\FOR{$k=1$ to $N$}
\STATE $\boldsymbol{\phi}_{k,n-1} = \sum_{l \in \mathcal{N}_k} a_{1,lk} \boldsymbol{\omega}_{l,n-1}$.
\STATE $\boldsymbol{\psi}_{k,n} = \boldsymbol{\phi}_{k,n-1} + \mu_k \sum_{l \in \mathcal{N}_k} c_{lk} \left| e_{l,n}\right|^{p-2} e_{l,n}\boldsymbol{u}_{l,n}$.
\STATE $\boldsymbol{\omega}_{k,n} = \sum_{l \in \mathcal{N}_k} a_{2,lk} \boldsymbol{\psi}_{l,n}$.
\ENDFOR
\ENDFOR
\end{algorithmic}
\caption{Diffusion LMP}\label{GDLMP}
\end{algorithm}

\section{Simulation Results}
\label{Sec5}
\begin{figure}[!ht]
\centering{\includegraphics[width=60mm]{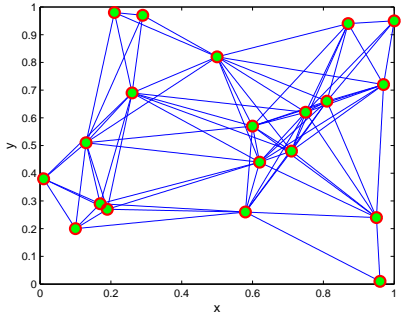}}
\caption{An ad hoc wireless sensor network with $N=20$ sensors, generated as a realization of the random geometric graph model on the unity square, with communication range $r = 0.5$.}
\label{fig1}
\end{figure}

We consider a connected ad hoc wireless sensor network composed of $N=20$ nodes. The network is generated as a realization of the random geometric graph model on the unity square, with communication range $r = 0.5$. The topology of the network is shown in Fig.\ref{fig1}.
The AR model can be rewritten as the following vector form for node $n$ at time $k$:
\begin{equation}
d_{k,n} = \mathbf{w}^T\mathbf{x}_{k,n} + v_{k,n}
\end{equation}
where $d_{k,n} = x_{k,n}$ is the desired signal, which is assumed to be white Gaussian process in the following simulations. $\mathbf{x}_{k,n} = [x_{k-1,n}, x_{k-2,n}, \cdots, x_{k-M,n}]^T$ is the input data and $\mathbf{w} = [w_1, w_2,\cdots, w_M]^T$ is the unknown parameters. $w_m$ is drawn randomly from the standard uniform distribution $\mathcal{U}(0,1)$.

Since the variance of alpha-stable process is infinite, instead of signal-to-noise ratio (SNR), we use generalized SNR (GSNR), which is defined in \cite{Tsakalides1995}. GSNR is the ratio of the signal power over the noise dispersion $\gamma$. All the following results are obtained by averaging over 10 independent Monte Carlo trials.

\begin{figure}[!ht]
\centering{\includegraphics[width=60mm]{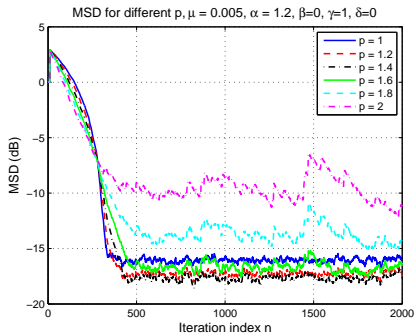}}
\caption{Transient network MSD of LMP for different $p$, step-size $\mu = 0.005$, $\alpha = 1.2$, $\beta=0$, $\gamma=1$, $\delta=0$. GSNR $= 20 dB$.}
\label{fig2}
\end{figure}

\begin{figure}[!ht]
\centering{\includegraphics[width=60mm]{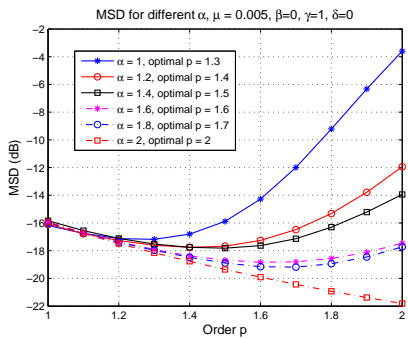}}
\caption{Steady network MSD of LMP as a function of order $p$ for different $\alpha$, step-size $\mu = 0.005$, $\beta=0$, $\gamma=1$, $\delta=0$. GSNR $= 20 dB$.}
\label{fig3}
\end{figure}

The transient and steady network MSD of LMP algorithms as a function of order p are given in Fig. \ref{fig2} and Fig. \ref{fig3}, respectively.  
From Fig. \ref{fig2}, we observe that, due to the impulsive noise, the LMP algorithm is unconverged for order $p = 2$  or $p$ close to 2. Meanwhile, the algorithm is converged when $p = 1$ or $p$ close to 1. 
From Fig. \ref{fig3}, we observe that the more impulsive of the noise (smaller $\alpha$), the larger MSD for $p=2$. The impulsive noise has a great influence on diffusion LMS algorithms. 
Furthermore, the lower MSD is obtained when $p$ is close to $\alpha$. Similar observations are obtained in \cite{Weng2005}, which is a single node based LMP algorithm for system identification.

\section{Conclusion}
We propose a diffusion least mean p-norm algorithm for distributed estimation in alpha stable noise environments. Compared with the diffusion LMS type algorithm, for impulsive noise with alpha-stable distribution, better performance is obtained for the diffusion LMP method. 
%
%
%

\end{document}